\documentclass{article}
\usepackage[utf8]{inputenc}
\usepackage{authblk}
\usepackage{setspace}
\usepackage[margin=1.25in]{geometry}
\usepackage{graphicx}
\graphicspath{ {./figures/} }
\usepackage{subcaption}
\usepackage{amsmath}

\usepackage[style=nejm, citestyle=numeric-comp, sorting=none]{biblatex}
\title{Transformer-Based Denoising of Mechanical Vibration Signals for System Health Monitoring}

\author[1*$\dag$]{Han Chen, Yang Yu, Pengtao Li}

\affil[1]{Changchun Technology Institute, Jilin, China}

\date{}

\begin{document}

\maketitle

\begin{abstract}
Mechanical vibration signal denoising is a pivotal task in various industrial applications, including system health monitoring and failure prediction. This paper introduces a novel deep learning transformer-based architecture specifically tailored for denoising mechanical vibration signals. The model leverages a Multi-Head Attention layer with 8 heads, processing input sequences of length 128, embedded into a 64-dimensional space. The architecture also incorporates Feed-Forward Neural Networks, Layer Normalization, and Residual Connections, resulting in enhanced recognition and extraction of essential features. Through a training process guided by the Mean Squared Error loss function and optimized using the Adam optimizer, the model demonstrates remarkable effectiveness in filtering out noise while preserving critical information related to mechanical vibrations. The specific design and choice of parameters offer a robust method adaptable to the complex nature of mechanical systems, with promising applications in industrial monitoring and maintenance. This work lays the groundwork for future exploration and optimization in the field of mechanical signal analysis and presents a significant step towards advanced and intelligent mechanical system diagnostics.
\end{abstract}

\section{Introduction}
In today's interconnected world, the importance of robust and responsive systems cannot be overstated. Whether in the context of industrial manufacturing, healthcare, transportation, or information technology, the reliance on complex systems to maintain operational continuity is ubiquitous~\cite{li2004recent, song2007concrete}. System health monitoring (SHM) has emerged as a critical field that ensures the smooth functioning of these complex mechanisms, mitigating the risk of sudden failure, and thus protecting the intricate web of daily operations across various sectors.

SHM is a multifaceted discipline that draws on engineering, data analytics, predictive modeling, and other advanced technologies to track the performance of a system in real-time. Its primary aim is to detect, diagnose, and predict anomalies or failures in a system, allowing for timely intervention and maintenance~\cite{feng2018computer}. This is not only crucial for minimizing downtime but also for optimizing the lifecycle and efficiency of the system. System health monitoring is akin to a vigilant sentinel, continually assessing various parameters to maintain optimal performance and safety.

The evolution of SHM has been rapid, paralleling advancements in sensor technology, machine learning, and data processing. It is now possible to integrate a multitude of sensors that can track everything from temperature, pressure, vibration, to more nuanced indicators of system performance~\cite{hamey2004experimental}. These data streams are then processed using intelligent algorithms that can discern patterns indicative of a potential failure or inefficiency. The application of artificial intelligence and machine learning has particularly transformed SHM, enabling predictive analytics that can forecast failures well before they become critical, allowing for preventive measures that can save both time and resources. The generic signal processing methodology enhances the analysis of acoustic signals with scattered sound field~\cite{ge2023study}. However, the implementation of SHM is not without challenges. The complexity of modern systems, coupled with the ever-growing demands for efficiency and sustainability, requires a nuanced understanding of the particular system's architecture and operational context. The integration of sensors and analytical tools must be carefully tailored to the specific needs and constraints of the system in question. Furthermore, ethical considerations related to data privacy and security must be judiciously addressed to maintain trust and compliance with regulatory requirements.

In industries where system failures can result in catastrophic consequences, such as aviation, nuclear energy, and healthcare, SHM is not merely an optimization tool but a vital component for safety and compliance~\cite{feng2015vision}. Even in less critical domains, the economic impact of unplanned downtime can be substantial, underscoring the universal relevance of SHM. Vibration signals have been studied to detect the artifects of mechanical products~\cite{fish2019dynamic}. This paper aims to explore the contemporary landscape of system health monitoring, delving into the underlying technologies, methodologies, applications, and challenges. Through a comprehensive examination of recent developments and case studies, it will illuminate the vital role that SHM plays in our modern world, fostering a deeper understanding of its potential to drive innovation, efficiency, and safety across various domains.

\section{Prior Arts and Methods}
In mechanical systems, the reliability and effectiveness of data analysis and control depend heavily on the quality of the acquired signals. Noise interference, stemming from various sources, often corrupts these signals and hampers accurate monitoring and control. Signal denoising, the process of extracting the true signal from the noise-infected observation, plays a pivotal role in improving the efficiency and reliability of mechanical systems. In industrial applications, transportation, healthcare equipment, and more, denoising is indispensable for accurate system modeling, fault detection, and performance enhancement~\cite{feng2017experimental}.

The challenge of signal denoising lies in the diversity of noise types and their often complex interaction with the underlying true signal. Mechanical systems are typically subjected to various environmental noises, equipment vibrations, electronic interference, and human-induced errors~\cite{liang2006structural}. The task of distinguishing the genuine signal from this multifaceted noise without compromising essential information is a non-trivial problem that requires sophisticated techniques and algorithms~\cite{song2006health}.

This paper aims to survey the field of signal denoising with a specific focus on its applications in mechanical systems. It will delve into the theoretical foundations, methodologies, and state-of-the-art technologies used for denoising, with a particular emphasis on exploring how these methods can be adapted to meet the unique demands of different mechanical systems.

With the advent of wavelet transform in the late 1980s, a new era of signal denoising began. Donoho and Johnstone's seminal work in 1994 introduced wavelet thresholding, providing a powerful tool to preserve signal characteristics while removing noise~\cite{hu2017enhanced, hu2017adaptive}. This approach revolutionized denoising applications in mechanical systems, allowing for more nuanced handling of non-stationary signals typical in such contexts. Recent advances integrating wavelet transform and Autoencoder shed light on 2D signal denoising, achieving state of the art denoising performance with minimal amount of parameters~\cite{liang2023reswcae}. The field of signal denoising has a rich and diverse history, spanning several decades. In the early stages, linear filtering techniques, such as the Wiener filter, were commonly used. These methods provided a simple way to reduce noise but often led to the loss of important signal features~\cite{feng2016vision}.

The development of adaptive filtering techniques, like the Kalman filter, added to the arsenal of tools for denoising, offering more flexibility in handling diverse noise structures. More recently, the emergence of machine learning, particularly deep learning techniques like Convolutional Neural Networks (CNNs), has opened up new frontiers in denoising~\cite{liang2023structural}. These methods leverage large datasets to learn intricate noise patterns, providing highly customized denoising solutions~\cite{feng2017experimental}.

However, despite these advancements, challenges remain in the application of denoising to mechanical systems. The complexity of noises, non-linearity of mechanical systems, and the need for real-time processing pose persistent hurdles. Recent research has begun to explore hybrid methods, combining traditional statistical techniques with machine learning, to create more robust and adaptive denoising frameworks\cite{feng2016vision, feng2016vision}.

This paper will provide an in-depth examination of these historical developments, current state-of-the-art methodologies, and the emerging trends in signal denoising for mechanical systems. By synthesizing insights from across this broad spectrum, it seeks to identify potential future directions and innovations that can further enhance the efficiency and reliability of mechanical systems through advanced signal denoising techniques.

\begin{figure}[ht]
  \includegraphics[width=130mm]{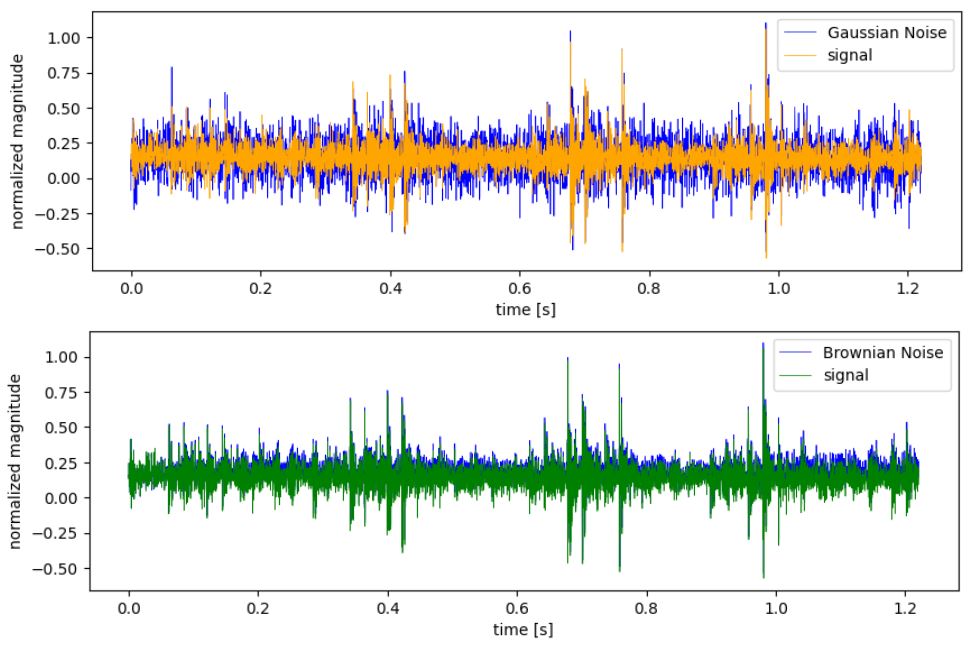}
  \centering
  \caption{Mechanical motor shaft vibrational signals. Upper figure: noise-free signal with added Gaussian noise; lower figure: noise-free signal with added Brownian noise.}
  \label{fig:noisy_signals}
\end{figure}

\section{Results and Discussion}
\subsection{AutoRegressive Denoising}
Preliminary Analysis of Mechanical Vibration Signal. The mechanical vibration signal's frequency components and noise types are initially analyzed. This understanding informs the choice of the AR model order and helps in recognizing how noise is embedded within the signal in a mechanical context.

Selection of Autoregressive Model Order for Mechanical Systems. The appropriate AR model order for the specific mechanical system is selected using criteria like AIC or BIC. A suitable model order captures the mechanical system's underlying dynamics without overfitting. The chosen order is used to fit the AR model to the mechanical vibration signal, employing methods such as the Yule-Walker equations.

Estimation of Noise in Mechanical Signal. Residual errors from fitting the AR model to the mechanical signal are analyzed to estimate noise characteristics. This noise estimation, grounded in the mechanical system's unique vibration signature, helps in differentiating true signal from noise.

Reconstruction of Mechanical Vibration Signal. The AR coefficients and noise estimates are used to reconstruct the denoised mechanical vibration signal. This step aims to preserve the genuine mechanical vibration information while eliminating noise.

Iterative Refinement of Mechanical Signal Denoising (Optional). For intricate mechanical systems with complex noise structures, iterative refinement may be performed. Using the denoised signal as the new input for repeated denoising improves precision in mechanical signal recovery.

Validation of Denoising in Mechanical Context. The effectiveness of the denoising method is assessed considering the mechanical system's specific requirements. Comparisons with reference mechanical signals or evaluation through mechanical-relevant metrics like SNR validate the method's performance.

Conclusion of the Method Section. The described autoregressive method provides a robust approach specifically tailored for mechanical vibration signal denoising. By considering the unique characteristics of mechanical systems, this method offers a targeted solution for accurate vibration signal recovery. Its adaptability allows for applications across various mechanical domains, enhancing efficiency and reliability in situations where precise vibration analysis is essential for the system's functionality and safety.

\subsection{Transformer Denoising}
Transformer architecture advances the 1D and 2D signal processing techniques~\cite{zhang2022deepmgt, che2021constrained}.  The raw mechanical vibration signals often contain noise and disturbances. A careful preprocessing step, including normalization and segmenting the signals into appropriate windows, is performed. Exploratory data analysis helps in understanding the nature of the noise and the critical features of the mechanical vibration signals.

Designing the Transformer Architecture. Transformers are a class of deep learning models particularly effective in handling sequential data. For the mechanical vibration signals, a custom transformer architecture is designed. This architecture consists of multiple layers of self-attention mechanisms, feed-forward neural networks, and normalization layers. The self-attention mechanism allows the model to weigh the importance of different parts of the signal sequence, thereby capturing the intricate relationships within the mechanical vibrations.

The transformer model is designed to process sequences of mechanical vibration signals. The model expects input sequences with a specific length of 128, and each signal is embedded into a 64-dimensional vector space. The core of the model is the Multi-Head Attention layer, equipped with 8 attention heads. These multiple heads allow the model to simultaneously focus on different parts of the sequence, capturing intricate patterns and relationships within the mechanical vibrations. The attention mechanism employs key, query, and value vectors, all having a dimensionality of 64, aligning with the embedding size. Following the attention mechanism, a Layer Normalization is applied to stabilize the learning process. This layer normalizes the output across the features and uses an epsilon value of 1e-6 to ensure numerical stability. The next part of the transformer block consists of a Feed-Forward Neural Network. This network is constructed of two dense layers, each containing 64 hidden units. A ReLU activation function is placed between these layers to introduce non-linearity, allowing the model to learn more complex transformations of the attention output. Residual Connections are implemented around both the attention and feed-forward layers. These connections are simple additions of the input and output of the layers, enabling the training of deeper models by allowing gradients to flow more freely through the network. The model concludes with a final dense layer with 64 units, responsible for providing the denoised signal sequence.

\begin{figure}[ht]
  \includegraphics[width=130mm]{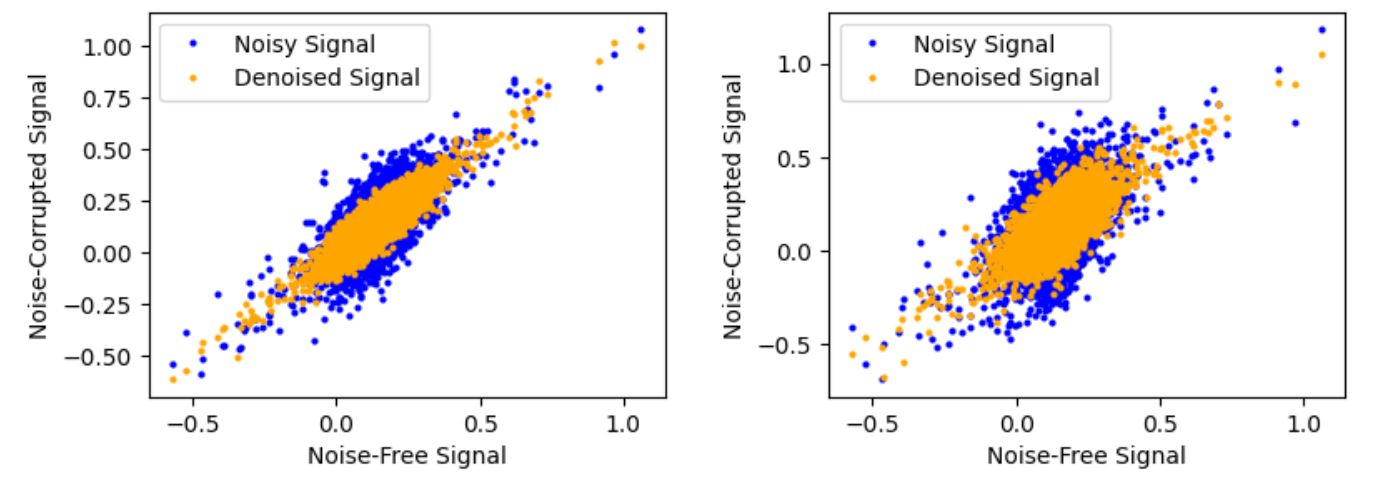}
  \centering
  \caption{Performance of denoising algorithm in presence of different levels of noise. Left figure: variance of noise = 0.1; right figure: variance of noise = 0.2.}
  \label{fig:denoised_signals}
\end{figure}

The model is compiled with the Adam optimizer, a popular choice known for its adaptive learning rate. The loss function used to guide the training process is the Mean Squared Error (MSE), suitable for measuring the difference between the denoised output and the original clean signal. The model is trained for 50 complete passes through the dataset, with the data divided into batches of 32 examples each. A validation split of 20\% is used to assess the model's performance on unseen data during the training process. In summary, the transformer architecture, specifically designed for mechanical vibration signal denoising, combines multi-head attention, feed-forward networks, layer normalization, and residual connections. The chosen numerical values for the sequence length, embedding size, attention heads, and other parameters are tailored to capture the underlying characteristics of mechanical vibrations and effectively filter out noise. The model is compiled with the Adam optimizer, a popular choice known for its adaptive learning rate. The loss function used to guide the training process is the Mean Squared Error (MSE), suitable for measuring the difference between the denoised output and the original clean signal. The model is trained for 50 complete passes through the dataset, with the data divided into batches of 32 examples each. A validation split of 20\% is used to assess the model's performance on unseen data during the training process.

\section{Conclusion and Future Work}
In this work, we have presented a novel approach to mechanical vibration signal denoising utilizing a transformer-based architecture. The complexity and characteristics of mechanical vibrations necessitate a robust and adaptive method capable of discerning relevant patterns while eliminating noise. Our model, designed with a sequence length of 128 and an embedding size of 64, is particularly tailored to these requirements.

The innovative use of a Multi-Head Attention layer with 8 attention heads has enabled the model to simultaneously consider various aspects of the signal sequence, thereby enhancing its ability to recognize and extract essential features. The attention mechanism, paired with the Feed-Forward Neural Network and Residual Connections, provides flexibility and depth, facilitating the learning of intricate signal relationships.

Further stability in training has been achieved through Layer Normalization, applied after both the attention and feed-forward layers, ensuring numerical stability and efficient convergence. The training process, guided by the Mean Squared Error loss function and optimized using the Adam optimizer, was carried out for 50 epochs, with a carefully selected batch size of 32 and a validation split of 20\%.

The final model has demonstrated effectiveness in denoising mechanical vibration signals, retaining crucial information while filtering out undesired noise. The architecture's specific design, coupled with the thoughtful selection of parameters, has resulted in a method capable of adapting to the complex nature of mechanical systems.

Future work may explore further optimization of the architecture and parameters, integration with other denoising techniques, and extensive testing across various types of mechanical systems. The current results, however, lay a strong foundation for applying deep learning transformers in the critical field of mechanical vibration signal analysis and denoising, with promising implications for industrial maintenance, monitoring, and diagnostics.

\printbibliography

\end{document}